# Surface-induced magnetic anisotropy for impurity spins in granular AuFe films

V. N. Gladilin[a,b], V. M. Fomin[a,b], J. T. Devreese[a,*]

[a]*Theoretische Fysica van de Vaste Stof, Universiteit Antwerpen (U.I.A.), Universiteitsplein 1, B-2610 Antwerpen (Belgium)*
[b]*Fizica Structurilor Multistratificate, Universitatea de Stat din Moldova, str. A. Mateevici 60, MD-2009 Chisinau (Moldova)*

**Abstract**

The theory of the surface-induced anisotropy is extended to the case of granular films of dilute magnetic alloys. Since the surface-induced blocking of a magnetic-impurity spin appears to be very sensitive to the specific polycrystalline structure, we speculate that the apparent discrepancy between the experimental results of different groups for the size dependence of the Kondo resistivity can be linked to different microstructure of the samples. We apply our model to calculate the magnetization of impurity spins in small AuFe grains and to interpret the experimental data on the anomalous Hall effect in thin Fe doped Au films.

**1. Introduction**

Recently, finite size effects in dilute magnetic alloys have been the subject of extensive experimental studies. While some experiments [1-3] revealed a considerable decrease of the logarithmic Kondo anomaly in the resistivity of thin films and narrow wires, other measurements [4] showed an almost constant Kondo anomaly for wires with width ranging down to 40 nm.

The size dependence of the Kondo effect in films and wires of dilute magnetic alloys can be linked to the surface-induced anisotropy of the magnetic-impurity spins [5-7]. In order to analyze the influence of a polycrystalline structure of the samples on the anisotropy effect, here we generalize the theory of the surface-induced magnetic anisotropy [5-7] to the case of mesoscopic grains. Based on the developed model, we analyze the size-dependent magnetization of impurities in those grains. The obtained results are applied to interpret the recent experimental data on the anomalous Hall effect in thin AuFe films [8].

**2. Surface-induced magnetic anisotropy in grains**

We have extended the approach, introduced in Ref. [7] for wires of dilute magnetic alloys, to the case of a brick-shaped grain with dimensions $a_x$, $a_y$, $a_z$. The Hamiltonian $H_{an}$, which describes the surface-induced anisotropy in an isolated grain, is obtained in the form

---

[*] e-mail: devreese@uia.ua.ac.be

$$H_{an} = \mathcal{A}\left[S_x^2 b(x,y,z) + S_y^2 b(y,x,z) + S_z^2 b(z,x,y)\right.$$
$$+ (S_x S_y + S_y S_x) c(x,y,z) + (S_x S_z + S_z S_x) c(x,z,y)$$
$$\left. + (S_y S_z + S_z S_y) c(y,z,x)\right]. \quad (1)$$

The $S_a$ $(a = x, y, z)$ are the operators for the components of the impurity spin $S$. The material dependent constant $\mathcal{A}$ should range between 0.01 and 1 eV for dilute AuFe alloys [6,7]. In a frame of reference with the origin at the center of the grain, the dependence of the anisotropy on the impurity position $(x,y,z)$ is described by the functions

$$b(\mathbf{a},\mathbf{b},\mathbf{g}) = \sum_{l,m,n=\pm 1} u(\mathbf{a}_l, \mathbf{b}_m, \mathbf{g}_n), \quad (2)$$

$$c(\mathbf{a},\mathbf{b},\mathbf{g}) = -\sum_{l,m,n=\pm 1} lm\, v(\mathbf{a}_l, \mathbf{b}_m, \mathbf{g}_n), \quad (3)$$

where

$$\mathbf{a}_l = |\mathbf{a} + l\mathbf{a}_a/2|, \quad (4)$$

$$u(p,q,t) = \frac{q}{p} v(p,q,t) + \frac{t}{p} v(p,t,q), \quad (5)$$

$$v(p,q,t) = \frac{1}{2pk_F \sqrt{p^2 + q^2}} \operatorname{arctg}\left(\frac{t}{\sqrt{p^2 + q^2}}\right), \quad (6)$$

$k_F$ is the Fermi wavenumber.

In grains, the presence of differently oriented surfaces leads to a rather intricate behavior of the impurity-spin anisotropy as compared to the case of a film considered in Refs. [5,6]. This is illustrated by Fig. 1, which shows the surface-induced splitting of the impurity-spin energy levels in AuFe grains. Note that, at any value of the constant $\mathcal{A}$, there exist specific regions within a grain, where the lowest state of an impurity spin is degenerate or quasi-degenerate. The impurities located in those regions can contribute to the Kondo effect even at relatively low temperatures. Such a partial (or even complete – depending on the position of an impurity and on the shape of a grain) cancellation of the anisotropy effect is a qualitatively new feature as compared to the case of a thin single-crystal AuFe film. In the latter case, the non-degenerate ground state is separated from other states by an energy interval $D > \mathcal{A}/d$ ($d$ is the film thickness) and, therefore, all the impurities with an integer spin become paramagnetic at temperatures $k_B T < D$ ($k_B$ is the Boltzmann constant) [5,6]. As seen from Fig. 1, the surface-induced magnetic anisotropy for impurity spins is very sensitive to the shape and size of grains. This implies that the apparent discrepancy between various experimental results [1-4] for the size dependence of the Kondo resistivity can be linked to different microstructures of the samples.

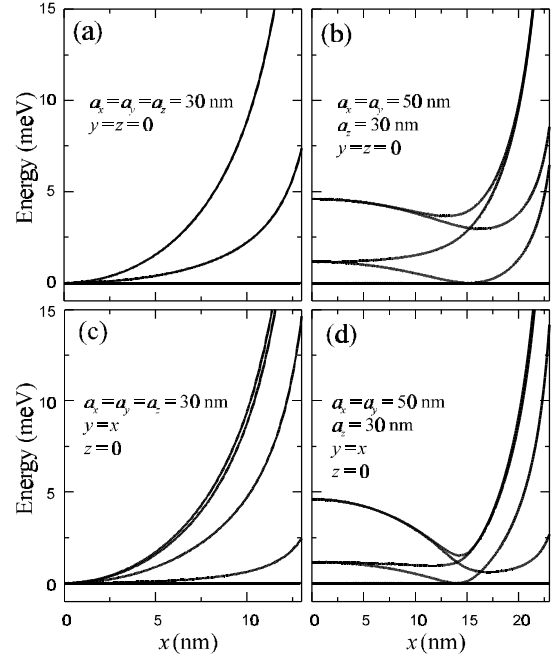

Fig. 1. Energy spectra of an impurity spin ($S=2$) for various impurity positions in a cubic grain [panels (a) and (c)] and in a flat grain [panels (b) and (d)] at $\mathcal{A} = 0.12$ eV. The energy of the lowest state of the impurity spin is taken to be zero.

## 3. Impurity-spin magnetization

In a magnetic field $\mathbf{B}$ parallel to the $z$-axis, the magnetization of an Fe spin is given by

$$\langle S_z \rangle = -\frac{1}{2m_B Z} \sum_{k=1}^{5} \exp\left(-\frac{E_k}{k_B T}\right) \frac{dE_k}{dB}, \quad (7)$$

where $m_B$ is the Bohr magneton, $T$ is the temperature, and $Z = \sum_{k=1}^{5} \exp(-E_k/k_B T)$. The index $k = 1,...,5$ labels the roots $E_k$ of the secular equation

$$|H_{S_z' S_z} - E d_{S_z' S_z}| = 0 \quad (S_z', S_z = -2,-1,0,1,2) \quad (8)$$

with the Hamiltonian $H = -2m_B S_z B + H_{an}$. The energy spectra of an impurity spin, which are shown in Fig. 2 as a function of magnetic field, appear to be qualitatively different for different positions of the impurity. Side by side with the energy spectra typical for magnetic impurities in a thin film subjected to a perpendicular magnetic field [see Fig. 2(a)], there are spectra similar to those in bulk [Fig. 2(b)] as well as spectra, which have no analogue in films or in bulk [Figs. 2(c) and 2(d)].

As shown in Fig. 3, the co-existence of impurity spins with substantially different energy spectra results in a strongly pronounced inhomogeneity of magnetization within a grain. In a flat grain, the differential magnetization at weak magnetic fields is significant only for impurities located near the grain edges parallel to the magnetic field, where the surface-induced magnetic anisotropy is appreciably weakened due to the competitive influence of mutually perpendicular surfaces. With increasing magnetic field, the magnetization of those impurities rapidly saturates, while the magnetic response of the impurities in the central region of the grain reveals itself only at relatively high magnetic field.

In Fig. 4 the calculated differential magnetization averaged over impurity positions within a grain, $[d\langle S_z \rangle / dB]_{gr}$, is shown as a function of the magnetic field for different lateral dimensions $a$ of the grains. In the limiting case of a single-crystal film ($a \to \infty$), the impurity spin states at $B = 0$ are known [5,6] to be the eigenstates of $S_z$ with the energy eigenvalues proportional to $S_z^2$. At low temperatures, only the state with $S_z = 0$ is populated. Hence, the impurity spin does not respond to a weak magnetic field. When increasing $B$, the energy level with $S_z = 1$ becomes lower then that with $S_z = 0$. This gives rise to the first peak in the differential magnetization as a function of magnetic field. The second peak appears when the state with $S_z = 2$ becomes the ground state. In grains, the presence of impurities with bulk-like behavior can result in a relatively high response at weak magnetic fields. For grains with lateral sizes $a$ larger than (but still comparable to) the height $a_z$, our model predicts that a minimum in $[d\langle S_z \rangle/dB]_{gr}$ can appear at moderate magnetic fields. For these grains the initial part of the calculated curves 'differential magnetization versus magnetic field' is similar to the curves 'differential Hall resistivity versus *B*', which have been recently obtained when measuring the anomalous Hall effect in granular AuFe films [8]. This striking similarity is illustrated by Fig. 5. Of course, the experimental samples contain grains of various shape and size and a detailed fitting of the experimental curve requires an averaging over an ensemble of grains. The results of such a fitting are reported elsewhere [8].

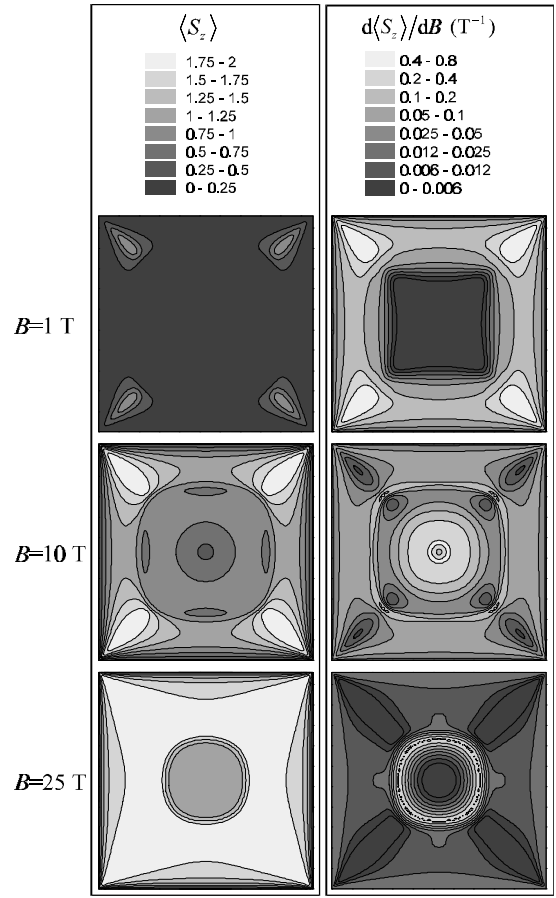

Fig. 3. Spatial distributions of the magnetization (l.h.s. panels) and of the differential magnetization (r.h.s. panels) for the cross-section $z = 0$ of a grain with $a_z = 30$ nm and $a = a_x = a_y = 50$ nm at $\mathcal{A} = 0.12$ eV and $T = 1$ K.

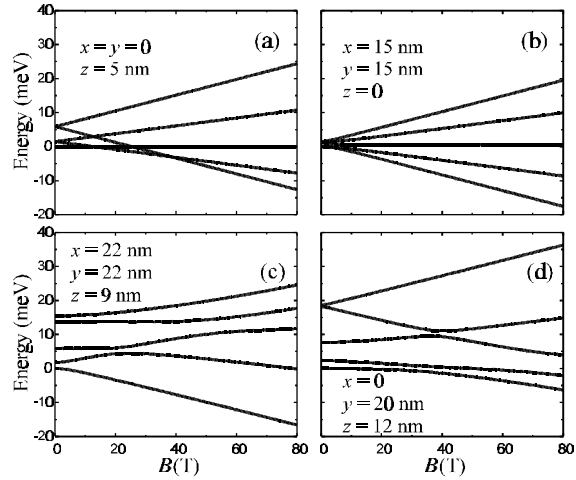

Fig. 2. Energy spectra of an impurity spin ($S=2$) as a function of magnetic field parallel to the $z$-axis are shown for various impurity positions in a grain with $a_z = 30$ nm and lateral size $a = a_x = a_y = 50$ nm at $\mathcal{A} = 0.12$ eV. The energy of the lowest state of the impurity spin at $B = 0$ is taken to be zero.

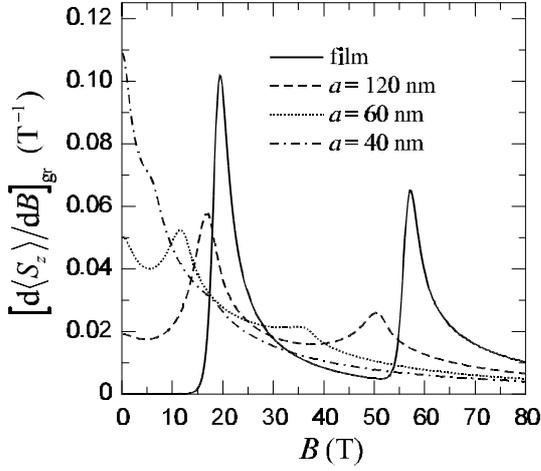

Fig. 4. Magnetic field dependence of the averaged differential magnetization for grains with height $a_z = 30$ nm and various lateral dimensions $a$ at $\mathcal{A} = 0.12$ eV and $T = 1$ K.

## 4. Conclusions

We have shown that, in small grains of dilute magnetic alloys, the competitive influence of differently oriented surfaces leads to a strong inhomogeneity of the surface-induced magnetic anisotropy for impurity spins. A high sensitivity of the anisotropy effect to the shape and size of grains implies that the apparent discrepancy between the experimental data of different groups [1-4] for the size dependence of the Kondo resistivity can be a result of a difference in the microstructure of the samples. Our model provides an explanation for the experimentally observed suppression of the anomalous Hall resistivity in thin polycrystalline AuFe films at low magnetic fields as well as for the appearance of a minimum in the differential Hall resistivity at higher fields [8].

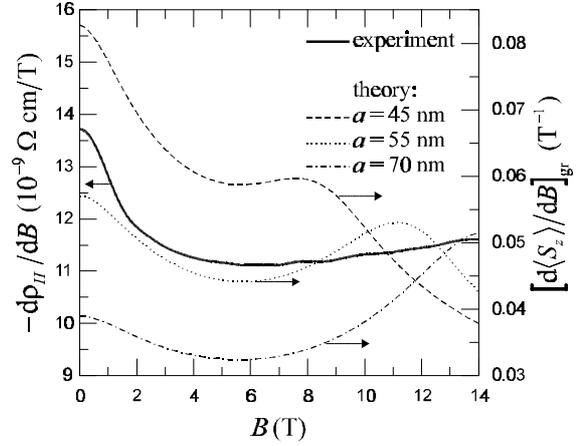

Fig. 5. Comparison between the measured differential Hall resistivity $-d\rho_H/dB$ for a 30 nm thick AuFe film with an Fe concentration 2 at.% [8] and the calculated averaged differential magnetization for grains with height $a_z = 30$ nm and various lateral dimensions $a$ at $\mathcal{A} = 0.12$ eV and $T = 1$ K.

## Acknowledgements

This work has been performed in collaboration with E. Seynaeve, K. Temst, F. G. Aliev, and C. Van Haesendonck (Katholieke Universiteit Leuven, Belgium). It has been supported by the I.U.A.P., GOA BOF UA 2000, F.W.O.-V. projects Nos. G.0287.95, 9.0193.97 and the W.O.G. WO.025.99N (Belgium).